\documentclass{aastex62}

\newcommand\aastex{AAS\TeX}

\received{November 15, 2018}
\revised{February 7, 2019}
\accepted{February 21, 2019}
\submitjournal{AJ}

\shorttitle{\aastex\ Observations of a near-Earth Object 2012 $\mathrm{TC_4}$}
\shortauthors{Urakawa et al.}
\usepackage{amsmath}

\begin{document}

\title{Shape and Rotational Motion Models for Tumbling and Monolithic Asteroid 2012 $\mathrm{TC_4}$:High Time Resolution Lightcurve with the Tomo-e Gozen Camera}

\correspondingauthor{Seitaro Urakawa}
\email{urakawa@spaceguard.or.jp}

\author[0000-0002-0786-7307]{Seitaro Urakawa}
\affil{Japan Spaceguard Association, Bisei Spaceguard Center 1716-3 Okura, Bisei, Ibara, Okayama 714-1411, Japan}

\author{Ryou Ohsawa}
\affiliation{Institute of Astronomy, Graduate School of Science, The University of Tokyo, 2-21-1 Osawa, Mitaka, Tokyo 181-0015, Japan}
%\collaboration{(AAS Journals Data Scientists collaboration)}

\author{Shigeyuki Sako}
\affiliation{Institute of Astronomy, Graduate School of Science, The University of Tokyo, 2-21-1 Osawa, Mitaka, Tokyo 181-0015, Japan}

\author{Shin-ichiro Okumura}
\affiliation{Japan Spaceguard Association, Bisei Spaceguard Center 1716-3 Okura, Bisei, Ibara, Okayama 714-1411, Japan}
%\nocollaboration

\author{Yuri Sakurai}
\affiliation{Department of Earth Science, Okayama University, 1-1-1 Kita-ku Tsushimanaka, Okayama 700-8530, Japan}
%\collaboration{(LaTeX collaboration)}

\author{Jun Takahashi}
\affiliation{Center for Astronomy, University of Hyogo 407-2 Nishigaichi, Sayo, Hyogo 679-5313, Japan}

\author{Kazuyoshi Imamura}
\affiliation{Anan Science Center, 8-1 Nagakawa Kamifukui Minami-Kawabuchi, Anan, Tokushima 779-1243, Japan}

\author{Hiroyuki Naito}
\affiliation{Nayoro Observatory, 157-1 Nisshin, Nayoro, Hokkaido 096-0066, Japan}

\author{Fumitake Watanabe}
\affiliation{Nayoro Observatory, 157-1 Nisshin, Nayoro, Hokkaido 096-0066, Japan}
\author{Ryoma Nagayoshi}
\affiliation{Nayoro Observatory, 157-1 Nisshin, Nayoro, Hokkaido 096-0066, Japan}
\author{Yasuhiko Murakami}
\affiliation{Nayoro Observatory, 157-1 Nisshin, Nayoro, Hokkaido 096-0066, Japan}

\author{Ryo Okazaki}
\affiliation{Asahikawa Campus, Hokkaido University of Education, 9 Hokumon, Asahikawa, Hokkaido 070-8621, Japan}

\author{Tomohiko Sekiguchi}
\affiliation{Asahikawa Campus, Hokkaido University of Education, 9 Hokumon, Asahikawa, Hokkaido 070-8621, Japan}

\author{Masateru Ishiguro}
\affiliation{Department of Physics and Astronomy, Seoul National University, 1 Gwanak-ro, Gwank-gu, Seoul 08826, Korea}

\author{Tatsuhiro Michikami}
\affiliation{Faculty of Engineering, Kindai University, Hiroshima Campus, 1 Takaya Umenobe, Higashi-Hiroshima, Hiroshima 739-2116, Japan}

\author{Makoto Yoshikawa}
\affiliation{Institute of Space and Astronautical Science, Japan Aerospace Exploration Agency, 3-1-1 Yoshinodai, Chuo-ku, Sagamihara, Kanagawa, 252-5210, Japan}

%% Note that the \and command from previous versions of AASTeX is now
%% depreciated in this version as it is no longer necessary. AASTeX 
%% automatically takes care of all commas and "and"s between authors names.

%% AASTeX 6.1 has the new \collaboration and \nocollaboration commands to
%% provide the collaboration status of a group of authors. These commands 
%% can be used either before or after the list of corresponding authors. The
%% argument for \collaboration is the collaboration identifier. Authors are
%% encouraged to surround collaboration identifiers with ()s. The 
%% \nocollaboration command takes no argument and exists to indicate that
%% the nearby authors are not part of surrounding collaborations.

%% Mark off the abstract in the ``abstract'' environment. 
\begin{abstract}

We present visible and near-infrared observations of a near-Earth object (NEO), 2012 $\mathrm{TC_4}$. The NEO 2012 $\mathrm{TC_4}$ approached close to the Earth at a distance of about 50,000 km in October 2017. This close approach provided a practical exercise for planetary defense. This apparition was also an appropriate opportunity to investigate 2012 $\mathrm{TC_4}$, which is a monolithic asteroid \citep{Polishook13}. We conducted the observation campaign of 2012 $\mathrm{TC_4}$ using six small- and medium-sized telescopes. The multiband photometry analysis showed that the taxonomic class of 2012 $\mathrm{TC_4}$ to be an X-type. In particular, we successfully obtained the high time resolution lightcurve of 2012 $\mathrm{TC_4}$ with the Tomo-e Gozen camera, which is the world's first wide-field CMOS camera, mounted on the 1.05 m Schmidt telescope at Kiso Observatory. The shape and rotational motion models of 2012 $\mathrm{TC_4}$ were derived from the lightcurve. When 2012 $\mathrm{TC_4}$ was assumed to be a triaxial ellipsoid, the rotational and precession periods were 8.47 $\pm$ 0.01 min and 12.25 $\pm$ 0.01 min, respectively, with the long axis mode. This indicates that 2012 $\mathrm{TC_4}$ is a tumbling and monolithic asteroid. The shape models showed that the plausible axial lengths to be 6.2 $\times$ 8.0 $\times$ 14.9~m or 3.3 $\times$ 8.0 $\times$ 14.3~m. The flattened and elongated shape indicates that 2012 $\mathrm{TC_4}$ is a fragment produced by a impact event. We also estimated the excitation timescale, which implied that the impact event happened within $\sim$3 $\times$ 10$^{5}$ yr and 2012 $\mathrm{TC_4}$ has a fresh surface.
\end{abstract}
%%12.245 minutes, and the precession period of 8.473
%% Keywords should appear after the \end{abstract} command. 
%% See the online documentation for the full list of available subject
%% keywords and the rules for their use.

%% From the front matter, we move on to the body of the paper.
%% Sections are demarcated by \section and \subsection, respectively.
%% Observe the use of the LaTeX \label
%% command after the \subsection to give a symbolic KEY to the
%% subsection for cross-referencing in a \ref command.
%% You can use LaTeX's \ref and \label commands to keep track of
%% cross-references to sections, equations, tables, and figures.
%% That way, if you change the order of any elements, LaTeX will
%% automatically renumber them.

%% We recommend that authors also use the natbib \citep
%% and \citet commands to identify citations.  The citations are
%% tied to the reference list via symbolic KEYs. The KEY corresponds
%% to the KEY in the \bibitem in the reference list below. 

\section{Introduction} \label{sec:intro}
``Planetary Defense" or ``Spaceguard" refers to a number of efforts against asteroid impact hazard. A spaceguard effort discovers near-Earth objects (NEOs), seeks their trajectory, and judges whether they will collide with the Earth. Representative NEO survey projects are Pan-STARRS (the Panoramic Survey Telescope and Rapid Response System, \citealp{Wainscoat14}; \citealp{Chambers16}), Catalina Sky Survey (\citealp{Larson98}; \citealp{Christensen14}), and NEOWISE \citep{Mainzer11a}. There are many other ground-based survey projects and future space plans including ATLAS \citep{Tonry11} and NEOCam \citep{Mainzer17}. In addition to survey observations, it is also important to reveal the physical properties of each NEO by obtaining information regarding the rotational period, rotational motion, shape, and taxonomic class. In the event of an impact hazard, such information can assist in the development countermeasure, such as a kinetic impactor \citep{Cheng18}. Moreover, fostering better understanding of NEOs helps to elucidate the planetary formation processes. Because NEOs have reflected the history of collision, destruction, and coalescence of small solar system bodies from the planet formation era. NEOs are also practically accessible object by spacecrafts. The physical properties that were estimated by the ground-based observations of asteroid (25143) Itokawa and asteroid (162173) Ryugu became essential information for the Hayabusa and Hayabusa-2 projects (\citealp{Kaasa03}; \citealp{Ostro05}; \citealp{Kim13}; \citealp{Ishiguro14}; \citealp{Muller17}). Furthermore, the technological progress brought on by explorations provides new prospects, such as manned explorations and resource collections of NEOs \citep{Abell16}. Exploration technology will also return to the spaceguard efforts as an impact avoidance technology. 

Hayabusa, Hayabusa-2, NEAR, and OSIRIS-Rex are representative NEO spacecrafts. The exploration of Itokawa by Hayabusa revealed that Itokawa was covered with numerous boulders and possessed rubble-pile structures due to weak gravity constraints \citep{Fujiwara06}. An asteroid that consists of a single boulder is sometimes called a monolithic asteroid. The physical properties of a monolithic asteroid, which could be the smallest unit constituting a rubble-pile asteroid, can provide clues to clarify the formation process of boulders in destructive collisions. An asteroid's rotational period is an important indicator to distinguish monolithic and rubble-pile asteroids. Although cohesive force might prevent the rotational breakup \citep{Rozitis14}, most asteroids rotating shorter than the period of 2.2 hr are considered to be monolithic asteroids because the fast rotation makes it difficult to keep the rubble-pile structure due to the strong centrifugal force \citep{Pravec00b}. Almost all monolithic asteroids are NEOs smaller than 200 m in diameter. Such NEOs are confirmed to be monolithic asteroids by measuring their rotational periods, immediately after being discovered by survey observations. Physical properties of monolithic asteroids, such as the taxonomic class and shape, are hardly determined, except for the rotational period and rough diameter. In order to estimate the taxonomic class, spectroscopic observations or multiband photometry is required. However, the small size and faintness of monolithic asteroids make it difficult to conduct spectroscopic observation, which demand sufficient brightness.  In the case of the multiband photometry, since the brightness of monolithic asteroids inevitably changes due to the fast rotation during the switching of the filter, we need to calibrate the brightness change appropriately. The calibration requires to obtain the accurate lightcurve data that cover the whole rotational phase of the monolithic asteroids. Otherwise, we need to calibrate the monolithic asteroids with a multiband simultaneous camera without the switching of the filter. Spectroscopic observations and multiband photometry are not carried out  immediately after the discovery of NEOs. Furthermore, the estimation of shape is required the enough amount of lightcurve data that are obtained by the observation of asteroid from various directions. To observationally deduce the taxonomic class and shape of the monolithic asteroids, the closest day-of-approach of the target asteroid to the Earth should be known in advance.  

The purpose of our study is to obtain the shape and rotational motion model of a NEO, 2012 $\mathrm{TC_4}$, from a high time resolution lightcurve. Furthermore, we deduce the taxonomic class of 2012 $\mathrm{TC_4}$ with the visible and near-infrared color indexes by multiband photometry. The NEO 2012 $\mathrm{TC_4}$ was discovered by the Pan-STARRS on October 4, 2012, and approached the Earth with a distance of $\sim$95,000~km on October 12, 2012. The rotational period and diameter were estimated to be 12.24 $\pm$ 0.06 min and 7-34~m, respectively \citep{Polishook13}. Therefore, 2012 $\mathrm{TC_4}$ is supposed to be a monolithic asteroid. However, the shape and taxonomic class were not identified. In addition, the lightcurve of 2012 $\mathrm{TC_4}$ was not fully explained by the period of 12.24~min. The NEO 2012 $\mathrm{TC_4}$ approached the Earth again on October 2017. The closest approach distance was $\sim$50,000~km on October 12, 2017. This apparition was an appropriate observation opportunity to investigate the physical properties of a small monolithic asteroid. In particular, we could use the Tomo-e Gozen camera (\citealp{Sako16}; \citealp{Sako18}), which was a low-noise, high-quantum-efficiency, and super wide-field CMOS mosaic camera. The quick and contiguous readout capability of the Tomo-e Gozen camera assisted in the observation of 2012 $\mathrm{TC_4}$, which is both fast-rotating and fast-moving. In this paper, we deal with the following. In Section 2, we describe the observations and their data reduction, with particular focus on the Tomo-e Gozen camera. In Section 3, we mention the results of taxonomic class, diameter, shape and rotational motion. In Section 4, we discuss the impact event that could have been happened on the parent object of 2012 $\mathrm{TC_4}$ and the excitation and damping timescales. Finally, we summarize the physical properties of 2012 $\mathrm{TC_4}$ and mention the significance of elucidating the physical properties of 10~m-sized NEOs.

\section{Observations and data reduction} \label{sec:obs}
\subsection{Observations}
We conducted the observation campaigns of 2012 $\mathrm{TC_4}$ with six small- and medium-sized telescopes from October 9 to October 11, 2017. Since 2012 $\mathrm{TC_4}$ moved to the dayside, we could not observe it on October 12, 2017, the day of the closest approach. The observational circumstances and states of 2012 $\mathrm{TC_4}$ are listed in Tables 1 and 2, respectively. The longest observation of this campaign was carried out using the Tomo-e Gozen camera mounted on a 1.05~m $f$/3.1 Schmidt telescope at Kiso Observatory. The Tomo-e Gozen is an extremely wide-field camera equipped with 84 CMOS sensors that consist of four modules with 21 CMOS sensors. The Tomo-e Gozen camera records an approximately 20 square degree area at a maximum frame rate of 2~Hz (= 0.5 s exposure). The field of view (FoV) for one CMOS sensor is 0.24 square degrees with a pixel resolution of 1.\arcsec2. The Tomo-e Gozen camera was not completed at the time 2012 $\mathrm{TC_4}$ approached the Earth, but a performance test of the Tomo-e Gozen camera was conducted using a single module with four CMOS sensors. The time control accuracy of the Tomo-e Gozen camera was around 1 s in the performance test. The quick and contiguous readout capability of the Tomo-e Gozen camera is suitable for the observation of a fast-rotating and fast-moving asteroid, such as 2012 $\mathrm{TC_4}$. The lightcurve of 2012 $\mathrm{TC_4}$ was obtained with high time resolution during the performance test. The exposure times were 10 s on October 9, 10 s and 5 s on October 10, and 2 s on October 11, 2017. To estimate the taxonomic class of 2012 $\mathrm{TC_4}$, spectroscopy was also conducted using the grism spectrometer of the Tomo-e Gozen camera with an exposure time of 5 s on October 11, 2017. However, the taxonomic class of 2012 $\mathrm{TC_4}$ was not estimated, because it was not possible to carry out the adequate wavelength calibration during the performance test. Despite this, the zeroth-order light in the grism spectroscopy was used as the lightcurve data. No filter was used in the observation at Kiso Observatory. 

The visible multiband photometry was performed using the 1.0~m $f$/3 telescope at Bisei Spaceguard Center (BSGC) on October 10, 2017. The multiband photometry data was also used as the lightcurve. The detector of the 1.0~m telescope consisted of four CCD chips with 4096 $\times$ 2048 pixels. We used one CCD chip to obtain as many images as possible by shortening the processing time. The FoV for one CCD chip is 0.65 square degrees with a pixel resolution of 1.\arcsec0. The multiband photometry data was obtained with Sloan Digital Sky Survey (SDSS) $g'$,$r'$,$i'$, and $z'$ filters. The filters were changed in the following sequence: 5 - 7 $g'$ images ${\rightarrow}$ 5 $r'$ images ${\rightarrow}$ 5 $i'$ images ${\rightarrow}$ 6 - 8 $z'$ images ${\rightarrow}$ 5 $i'$ images ${\rightarrow}$ 5 $r'$ images ${\rightarrow}$ 5 - 7  $g'$ images. We repeated this sequence three times. All images were obtained with an exposure time of 120 s for each filter in the no-binning mode. 

The near-infrared multiband photometry was carried out using the Nishiharima Infrared Camera (NIC) \citep{Takahashi14} mounted at the Cassegrain focus ($f$/12) of the 2.0~m Nayuta telescope at Nishi-Harima Astronomical Observatory. The FoV of the NIC is 2.\arcmin73 $\times$ 2.\arcmin73.  Since the NIC is a near-infrared three-band ($J$, $H$, and $K_{s}$) simultaneous camera, the color index of 2012 $\mathrm{TC_4}$ could be investigated without calibrating the change in rotational brightness. Only seven images were obtained, due to the poor weather on October 10, 2017, with an exposure time of 120 s.

In addition, we observed 2012 $\mathrm{TC_4}$ on October 10, 2017, using two 0.4~m $f/10$ telescopes equipped with SBIG STL-1001E CCD (1024 $\times$ 1024 pixels) at Nayoro Observatory. The FoV of each telescope was around  22\arcmin $\times$ 22\arcmin. The goal was to obtain the color index in the visible wavelength region by simultaneously imaging with each telescope, using an IDAS R filter and a Johnson V filter. The exposure time for each telescope was 30 s. The color index could not be estimated due to the poor S/N; however the photometric data helped complement the phase of the lightcurve. 

Finally, the 1.13~m $f$/9.7 telescope at Anan Science Center provided the photometric data for lightcurve on October 11, 2017. The detector and FoV were the SBIG STX-16803 CCD (4096 $\times$ 4096 pixels) and 11.\arcmin5  $\times$ 11.\arcmin5, respectively. The photometry was conducted with an exposure time of 6 s by the 2 $\times$ 2 binning mode without the use of a filter.

\subsection{Data Reduction for Lightcurve}
All images were bias and flat-field corrected. The observational time was corrected using the light-travel time from 2012 $\mathrm{TC_4}$ to the observatory site. To calibrate the magnitude fluctuations due to the change of atmospheric conditions, a relative photometry was conducted using reference stars imaged in the same frame as 2012 $\mathrm{TC_4}$:

\begin{equation}
F^{i}_{c}(t) = F^{i}_{0}(t) - \overline{F^{i}_{r}(t)},
\end{equation}
where $F^{i}_{c}(t)$ is the calibrated lightcurve of 2012 $\mathrm{TC_4}$ under the $i$th observational condition, namely, each observatory, each observation day, and each filter. $F^{i}_{0}(t)$ is the raw magnitude of 2012 $\mathrm{TC_4}$; $\overline{F^{i}_{r}(t)}$ is the average raw magnitude of the reference stars and represents the change of atmospheric conditions; $t$ is the observational time. There were 20-60 and three reference stars for Kiso Observatory and BSGC observations, respectively. One reference star was applied for the observations at Nayoro Observatory and Anan Science Center. Next, offset magnitudes $\Delta{F}^{i}$ were calculated to adjust the lightcurve of each observational condition. Here, we notes that the differences of the phase angle and the distances of 2012 $\mathrm{TC_4}$ among the observational conditions are not regarded in the relative photometry. The offset magnitudes $\Delta{F}^{i}$ was estimated as the difference of the average magnitude $\overline{F^{i}_{c}}$ and the standard average magnitude $\overline{F^{kiso10}_{c}}$, obtained on October 10, 2017, at Kiso Observatory:
\begin{equation}
\Delta{F}^{i} = \overline{F^{i}_{c}} - \overline{F^{kiso10}_{c}},
\end{equation}
where the standard average magnitude $\overline{F^{kiso10}_{c}}$ was estimated by comparing with the SDSS $g'$ magnitude of reference stars. The standard average magnitude was 17.575 mag. Since the Tomo-e Gozen was not equipped with the SDSS $g'$ filter, the standard average magnitude could be affected by a constant offset due to possible imperfect color correction. The constant offset was, however, estimated to be up to $\sim$0.1 mag, which has little impact on the following discussion. The calibration process above may introduce some systematic error in $\Delta{F}^{i}$, since each observation covered the different phase of the lightcurve and the average brightness should be different in the observational conditions. However, we presume that the systematic errors were sufficiently small, since the observation time of each observational condition was enough long compared with the rotational period as will be described later. We can safely use the average magnitude $\overline{F^{i}_{c}}$ to adjust the lightcurves obtained in the different observational conditions. Finally, the lightcurve of 2012 $\mathrm{TC_4}$, $F(t)$ could be described as   

\begin{equation}
{F(t)} = F^{i}_{c}(t)  + \Delta{F}^{i}.
\end{equation}

\subsection{Data Reduction for Multiband Photometry}
The multiband photometry in the visible wavelength region was conducted in BSGC. We measured the flux of 17 standard stars from the SDSS Data Release 12 \citep{Alam15}, whose stars were imaged simultaneously in the same frame as 2012 $\mathrm{TC_4}$. These objects have magnitudes of about 14-16 mag in the $g'$-band and classification code 1 (= primary), quality flag 3 (= good), and object class 6 (= star). The apparent magnitude of 2012 $\mathrm{TC_4}$ was derived using the conversion factors that were evaluated from the 17 standard stars. Since 2012 $\mathrm{TC_4}$ is fast-rotating, its brightness inevitably changes during the filter switch. We defined the time of recording the first $g'$ images as a standard time (JD = 2458036.9707319), and then we calibrated an amount of brightness change for the standard time. The amount of brightness change was estimated by the fitting curve of the lightcurve as will be described in the following chapter. The multiband photometry in the near-infrared wavelength region was conducted at the Nishi-Harima Observatory. To increase the photometric accuracy, the seven obtained images were stacked with the median. The photometric standard star was one 2MASS catalog star (source designation: 22552054-0349375, $J$ = 13.586 $\pm$ 0.027, $H$ = 13.003 $\pm$ 0.029, $K_{s}$ = 12.926 $\pm$ 0.033) that was imaged simultaneously with 2012 $\mathrm{TC_4}$ in two of the seven frames.

\section{Results}
\subsection{Lightcurve}
Assuming a double-peaked lightcurve, we carried out a periodicity analysis based on the Lomb-Scargle periodogram (\citealp{Lomb76}; \citealp{Scargle82}). We had a possibility to evaluate the inaccurate rotational period, due to the change in the geometric relationship between the Earth,  2012 $\mathrm{TC_4}$, and Sun for a few days. To avoid this, only the stable data obtained on October 10, 2017, at Kiso Observatory were used in the periodic analysis. The power spectrum from the periodogram showed a first period of 12.25 $\pm$ 0.01~min and a second period of 8.47 $\pm$ 0.01~min (Figure 1). The result was consistent with other observational results (\citealp{Sonka17}; \citealp{Warner18a}; \citealp{Tan18})\footnote{In addition,``The 2012 $\mathrm{TC_4}$ Observing Campaign." http://2012tc4.astro.umd.edu/index.shtml}. The appearance of two fast-rotating periods shows that 2012 $\mathrm{TC_4}$ is a tumbling and monolithic asteroid. Substituting 12.25~min and 8.47~min into $P_{1}$ and $P_{2}$, respectively, the period ratio $P_{1}$:$P_{2}$ became approximately 13:9. In order to fit a curve to the lightcurve of a tumbling asteroid,``$Combined$ $Period$" $P_{c}$ was defined by 
\begin{equation}
P_{c} = \frac{9P_{1} + 13P_{2}}{2}.
\end{equation}
The same surface of 2012 $\mathrm{TC_4}$ faces the observer every $P_{c}$ of 110.18~min. Next, we made the folded lightcurve with period of $P_{c}$ for each day. The common specific features appear in the different phases of the folded lightcurve for each day. The folded lightcurve covering three observation days was obtained by matching the specific features in the phase (Figure 2). The obtained lightcurve was fitted to the two-dimensional Fourier series \citep{Pravec05}:
\begin{eqnarray}
%%F(\psi(t),\phi(t))
%%\nonumber F(\psi(t),\phi(t)) &=& F^{m}(t)\\
\nonumber  F^{m}(t)                 & =& C_{0}+\sum^{m}_{j=1}\left[C_{j0}\cos\frac{2{\pi}j}{P_{1}}t+S_{j0}\sin\frac{2{\pi}j}{P_{1}}t\right]\\
\nonumber            &+& \sum^{m}_{k=1}\sum^{m}_{j=-m}\left[C_{jk}\cos\left(\frac{2{\pi}j}{P_{1}}+\frac{2{\pi}k}{P_{2}}\right)t\right.\\
            &+& \left. S_{jk}\sin\left(\frac{2{\pi}j}{P_{1}}+\frac{2{\pi}k}{P_{2}}\right)t\right],
\end{eqnarray}
where $m$ is the order; $C_{0}$ is the mean reduced light flux; $C_{jk}$ and $S_{jk}$ are the Fourier coefficients for the linear combination of the two frequency $P^{-1}_{1}$ and $P^{-1}_{2}$, respectively; and $t$ is the time. Substituting $m$ = 4, $P_{1}$ = 12.25~min, and $P_{2}$ = 8.47~min for 2012 $\mathrm{TC_4}$, a fitting curve was obtained, as shown by the blue lines in Figure 2. The $C_{0}$ value was 17.578 mag. The brightness was around the same with the standard average magnitude of 17.575 mag described in Section 2.2. This indicates quantitatively that the offset error obtained in Eq. (2) is enough small for the purpose of obtaining the fitting curve. The top of Figure 2 shows that the obtained data can cover almost all phases in the lightcurve. The second top of Figure 2 indicates that the data of multiband photometry on BSGC is distributed evenly to the phase of the lightcurve. This means that the precise color index can be evaluated by stacking the data for each filter, assuming that 2012 $\mathrm{TC_4}$ has a homogeneous surface. The graph legend ``Kiso 11. Oct/1st" in the bottom of Figure 2 shows the result of zeroth-order photometry by grism spectroscopy. Thus, the photometric accuracy is slightly worse than the result of photometry in the graph legend ``Kiso 11. Oct/2nd". A precise and high time resolution lightcurve was successfully obtained in the graph legend ``Kiso 11. Oct/2nd" by taking the advantage of a unique feature of the Tomo-e Gozen camera. The high time resolution lighcurve contributes to the drawing of a precise fitting curve. 

\subsection{Taxonomic Class and Diameter}
The taxonomic class in the visible wavelength region is investigated in terms of a reflectance color gradient and a log reflectance spectrum \citep{Carvano10}. The reflectance color gradient and log reflectance spectrum are deduced from the color indexes of 2012 $\mathrm{TC_4}$. Although the adequate wavelength calibration could not be carried out in grism spectroscopy at Kiso Observatory, the spectrum feature did not show the time variation. This indicated that 2012 $\mathrm{TC_4}$ had a homogeneous surface and the color indexes did not also show the time variation. Assuming the homogeneous surface of 2012 $\mathrm{TC_4}$, the photometric accuracy could be increased by averaging the flux for each filter. The color indexes were $g' - r'$ =  0.479 ${\pm}$ 0.031,  $r' - i'$ = 0.187 ${\pm}$ 0.023, and $i' - z'$ = 0.035 ${\pm}$ 0.036, respectively. The reflectance color is defined as 
\begin{equation}
 C_{\lambda_{j}}= -2.5 (log_{10}R_{\lambda_{j}} - log_{10}R_{\lambda_{ref}}),
\end{equation}
where $C_{\lambda_{j}}$ and $R_{\lambda_{j}}$ are the reflectance color and the reflectance at a given wavelength; $R_{\lambda_{ref}}$ is the reflectance at the reference wavelength; and the subscript $j$ specifies the wavelength. The wavelengths of $j$ = 1, 2, 3, and 4 correspond to the central wavelengths of the SDSS $g'$(0.477 $\mu$m), $r'$(0.623 $\mu$m), $i'$(0.763 $\mu$m), and $z'$(0.913 $\mu$m) filters, respectively. When we use the $g'$ filter as the reference, the reflectance colors of the $r'$ filter is calculated from the color index as
\begin{equation}
 C_{r}= (r' - g') - C_{\odot_{rg}},
\end{equation} 
where $C_{\odot_{rg}}$ is the $r'$ - $g'$ color of the Sun. We adopted the solar colors $C_{\odot_{rg}}$, $C_{\odot_{ig}}$, and $C_{\odot_{zg}}$ \citep{Ive01}. The reflectance colors of the other filters were calculated in the same manner. The reflectance color gradient is defined as 
\begin{equation}
 {\gamma_{j}}= -0.4\frac{C_{\lambda_{j+1}} - C_{\lambda_{j}}}{\lambda_{j+1} - {\lambda_{j}}}.
\end{equation}
We deduced the reflectance color gradients of $\gamma_{g}$ = 0.079 $\pm$ 0.038, $\gamma_{r}$ = 0.249 $\pm$ 0.043, and $\gamma_{i}$ = -0.013 $\pm$ 0.052. Figure 3 shows the reflectance color gradients of 2012 $\mathrm{TC_4}$ and asteroids of major taxonomic classes. The rectangles in Figure 3 indicates the range of reflectance color gradients of C, X, D, L, S, A, Q, O, and V-type asteroids in the SDSS Moving Object Catalog (SDSS-MOC). The top, middle and bottom figures correspond to the $\gamma_{g}$, $\gamma_{r}$, and $\gamma_{i}$, respectively. The thick horizontal lines are the average reflectance color gradients of 2012 $\mathrm{TC_4}$. The reflectance color gradients of 2012 $\mathrm{TC_4}$ are consistent with the range of X-type asteroids. The log reflectance for 2012 $\mathrm{TC_4}$ was normalized at the $g'$ filter. The normalized log reflectance of the $r'$, $i'$, and $z'$ were 1.01 $\pm$ 0.01, 1.05 $\pm$ 0.01, and 1.04 $\pm$ 0.02, respectively. Figure 4 shows the log reflectance spectra of 2012 $\mathrm{TC_4}$ and the asteroids of the X-, S-, C-, and L-types. The log reflectance spectrum of 2012 $\mathrm{TC_4}$ is similar to that of the X-type. Both observational results indicate that the taxonomic classes of 2012 $\mathrm{TC_4}$ in the visible wavelength region is the X-type. Moreover, the color indexes in the near-infrared wavelength region were $J - H$ =  0.226 ${\pm}$ 0.041 and $H - K_{s}$ = 0.034 ${\pm}$ 0.045. These values were included in the range of C-complex ($J - H$ =  0.28 ${\pm}$ 0.08,  $H - K_{s}$ = 0.11 ${\pm}$ 0.08), S-complex ($J - H$ =  0.37 ${\pm}$ 0.12,  $H - K_{s}$ = 0.04 ${\pm}$ 0.08), and X-complex ($J - H$ =  0.31 ${\pm}$ 0.12,  $H - K_{s}$ = 0.14 ${\pm}$ 0.07) \citep{Popescu16}. Therefore, the taxonomic class of 2012 $\mathrm{TC_4}$ was concluded to be an X-type. The color indexes are summarized in Table 3.

We estimate the absolute magnitude $H_{V}$ and effective diameter of 2012 $\mathrm{TC_4}$. The average apparent $r'$ magnitude of 2012 $\mathrm{TC_4}$ on October 10, 2017, at BSGC, was deduced to be 17.129 $\pm$ 0.017 mag. The apparent  $V$ magnitude is described in the following form \citep{Fukugita96}:

 \begin{equation}
 V=r'-0.11+0.49\left(\frac{(g'-r')+0.23}{1.05}\right).
 \end{equation}
Here, for our photometric precision requirements, the difference between the AB magnitude and Vega magnitude in the $V$ band is negligible. The reduced magnitude at the phase angle, $\alpha$, is expressed as $H(\alpha) = V-5\log_{10}(R\Delta)$, where $R$ and $\Delta$ are the heliocentric and geocentric distances in au, respectively. The absolute magnitude is expressed as a so-called $H$-$G$ function \citep{Bow89}: 

\begin{equation}
H_{V}=H(\alpha)+2.5\log_{10}[(1-G)\Phi_{1}(\alpha)+G\Phi_{2}(\alpha)],
\end{equation}
where $G$ is the slope parameter dependent on the asteroid's taxonomy. When we apply $G$ = 0.20 $\pm$ 0.09 for X-types \citep{Veres15}, $H_{V}$ becomes 28.54 $\pm$ 0.03 mag. An effective diameter of asteroids $D$ (in kilometer) is described as

\begin{equation}
D = 1329\times10^{-H_{V}/5}p_{V}^{-1/2},
\end{equation}
where $p_{V}$ is the geometric albedo. Assuming an albedo of 0.098 $\pm$ 0.081 for the X-type \citep{Usui13}, the effective diameter and range were found to be 8 m and 6 m $<$ $D$ $<$ 20 m, respectively. Since \cite{Mainzer11b} also showed the albedo of 0.099 $\pm$ 0.161 for Tholen X-complex class and the albedo of 0.111 $\pm$ 0.143 for Bus-DeMeo X-complex class, the assumption of $\sim$0.1 for the X-type albedo was reasonable. We should note, however, the X-complex includes the E-, M-, and P-types, whose albedos are 0.454 $\pm$ 0.119, 0.169 $\pm$ 0.044, and 0.063 $\pm$ 0.017, respectively \citep{Usui13}. The estimated diameter can be affected by the uncertainty of the albedo among the X-complex asteroids. 

\subsection{Shape and Rotational Motion}
The period analysis revealed that 2012 $\mathrm{TC_4}$ is a tumbling asteroid with a rotational period and precession period. However, period analysis alone cannot conclude whether a given period, $P_{1}$ or $P_{2}$, corresponds to the rotational period or precession period. Thus, we make the shape and rotational motion models of 2012 $\mathrm{TC_4}$, which is recognized as a force-free asymmetric rigid body, from the dynamic analytical solution. Previous studies have described the equations of motion for a force-free asymmetric rigid body (\citealp{Samarasinha91}; \citealp{Kaasa01a}). The main equations used in this study are detailed in the Appendix. The shape and rotational motion models of 2012 $\mathrm{TC_4}$ were made by substituting the observational result into the equations of this subsection and the Appendix. However, it should be noted that the shape and rotational motion models are representative examples, not unique solutions. Here, we define $L_{s}$ (short axis length), $L_{i}$ (intermediate axis length), and $L_{l}$ (long axis length) when 2012 $\mathrm{TC_4}$ is a triaxial ellipsoid body. The axes satisfy the relationship $L_{s} < L_{i} < L_{l}$. The rotational motions of asteroids are categorized into long axis modes (LAM) and short axis modes (SAM). The body of LAM rotates completely around the long axis ($\psi$ in the Appendix) and oscillates around the short axis ($\phi$ in the Appendix), as seen by an external observer. On the other hand, the body of SAM oscillates around the long axis ($\psi$ in the Appendix) and rotates fully around the short axis ($\phi$ in the Appendix), as seen by an external observer. The shape and rotational motion models of 2012 $\mathrm{TC_4}$ were made for LAM and SAM, respectively.

First, the LAM models were made. The relation between the lightcurve amplitude and phase angle is shown as follows: 

\begin{equation}
A(0) = \frac{A(\alpha)}{1+c\alpha},
\end{equation}
where $A(\alpha)$ is the lightcurve amplitude at the phase angle, $\alpha^{\circ}$, and $c$ is the photometric phase slope coefficient. Since the X-type of 2012 $\mathrm{TC_4}$ includes the E-type, M-type, and P-type, 0.03 of the M-type \citep{Zap90} was adopted as the $c$ value. Assuming that the light-scattering cross-section of 2012 $\mathrm{TC_4}$ is projected onto the plane of the sky, the lightcurve amplitude is described through the lower limit to the true cross-section ratio of the body as
\begin{equation}
A(0) = 2.5\log_{10}\left(\frac{S_{max}}{S_{min}}\right),
\end{equation}
where $S_{max}$ and $S_{min}$ are the maximum and minimum light-scattering cross-sections, respectively. The maximum amplitude of 2012 $\mathrm{TC_4}$ is 1.434~mag and appears at a phase around 0.2 in Figure 2 when the phase angle is $39^{\circ}$. Therefore, the relationship $L_{l} = 2.40L_{s}$ was obtained, assuming $S_{max}$ = $\pi$$L_{l}L_{i}$ and $S_{min}$ = $\pi$$L_{i}L_{s}$. Alternatively, the relationship $L_{i} = 2.40L_{s}$ could be obtained from $S_{max}$ = $\pi$$L_{l}L_{i}$ and $S_{min}$ = $\pi$$L_{l}L_{s}$, when 2012 $\mathrm{TC_4}$ almost simply rotates around the long axis and an observer sees 2012 $\mathrm{TC_4}$ from the vertical direction for the total rotational angular momentum vector. As described above, period analysis alone is insufficient to determine whether $P_{1}$ or $P_{2}$ corresponds to $P_{\psi}$ (period of $\psi$) or $P_{\phi}$ 
 (period of $\phi$) in the Appendix. Thus, there were four cases for the LAM models, whose combinations were $L_{l} = 2.40L_{s}$ or $L_{i} = 2.40L_{s}$ for $P_{\psi}$ = 12.25~min and $P_{\phi}$ = 8.47~min or $P_{\psi}$ = 8.47~min and $P_{\phi}$ = 12.25~min. Substituting the axial ratios and periods to Eq. (A11), the following limits of $L_{s}$, $L_{i}$, and $L_{l}$ could be taken for four cases:\\
$\bullet$ Case 1: $L_{l} = 2.40L_{s}$, $P_{\psi}$ = 12.25~min, $P_{\phi}$ = 8.47~min, $L_{i} \leq 1.88L_{s}$\\
$\bullet$ Case 2: $L_{i} = 2.40L_{s}$, $P_{\psi}$ = 12.25~min, $P_{\phi}$ = 8.47~min, $L_{l} \geq 2.97L_{s}$\\
$\bullet$ Case 3: $L_{l} = 2.40L_{s}$, $P_{\psi}$ = 8.47~min, $P_{\phi}$ = 12.25~min, $L_{i} \leq 1.39L_{s}$\\
$\bullet$ Case 4: $L_{i} = 2.40L_{s}$, $P_{\psi}$ = 8.47~min, $P_{\phi}$ = 12.25~min, $L_{l} \geq 3.69L_{s}$\\
The combination of the average rotational velocities were $\overline{\dot\phi}$ $\sim$ 42.5~$deg{\cdot}min^{-1}$ and $\overline{\dot\psi}$ $\sim$ 29.4~$deg{\cdot}min^{-1}$ or $\overline{\dot\phi}$ $\sim$ 29.4~$deg{\cdot}min^{-1}$ and $\overline{\dot\psi}$ $\sim$ 42.5~$deg{\cdot}min^{-1}$. Moreover, we applied $L_{i}$ = 8~m as the effective diameter of  2012 $\mathrm{TC_4}$. In addition, the moments of inertia of Eq. (A4) were given using the total rotational angular momentum $M$ and total rotational kinetic energy $E$ as follows:
\begin{equation}
\frac{M^2}{2E} = \frac{(nI_{l} + I_{i})}{n+1}\quad or \quad \mbox{ }\frac{(I_{l} + mI_{i})}{m+1},
\end{equation}
where $n$ is an integer from 1 to 9 and $m$ is an integer from 2 to 9. The tumbling status were roughly described by changing the integer $n$ and $m$. The combination of the $n$, $m$, and axial lengths that satisfies the observed velocities $\overline{\dot\phi}$ and $\overline{\dot\psi}$ was sought. The axial lengths were scanned in steps of 0.1 m for each case. The results were the following combinations:\\
$\bullet$ Case 1': ($L_{s}$, $L_{i}$, $L_{l}$) = (7.5 m, 8.0 m, 18.0 m), $M^2/2E$ = $(I_{l} + 3I_{i})/4$\\
$\bullet$ Case 3': ($L_{s}$, $L_{i}$, $L_{l}$) = (6.2 m, 8.0 m, 14.9 m), $M^2/2E$ = $(8I_{l} + I_{i})/9$\\
$\bullet$ Case 4': ($L_{s}$, $L_{i}$, $L_{l}$) = (3.3 m, 8.0 m, 14.3 m), $M^2/2E$ = $(2I_{l} + I_{i})/3$\\
There was no solution that satisfied the observed velocities $\overline{\dot\phi}$ and $\overline{\dot\psi}$ for ``Case 2". 

Next, the SAM models were made. The relationship of $L_{l} = 2.40L_{s}$ was obtained, assuming $S_{max}$ = $\pi$$L_{l}L_{i}$ and $S_{min}$ = $\pi$$L_{i}L_{s}$. Alternatively, the relationship of $L_{l} = 2.40L_{i}$ could be obtained from $S_{max}$ = $\pi$$L_{l}L_{s}$ and $S_{min}$ = $\pi$$L_{i}L_{s}$, in the case that 2012 $\mathrm{TC_4}$ almost simply rotates around the short axis and an observer sees 2012 $\mathrm{TC_4}$ from the vertical direction for the total rotational angular momentum vector. The combination of the rotational period and precession period is limited to be $P_{\psi}$ = 12.25~min and $P_{\phi}$ = 8.47~min by Eq. (A20). Substituting the axial ratios and periods into Eq. (A19), the following limit of $L_{s}$, $L_{i}$, and $L_{l}$ could be taken for two cases:\\
$\bullet$ Case 5: $L_{l} = 2.40L_{s}$, $P_{\psi}$ = 12.25~min, $P_{\phi}$ = 8.47~min, $L_{i} \geq 2.29L_{s}$\\
$\bullet$ Case 6: $L_{l} = 2.40L_{i}$, $P_{\psi}$ = 12.25~min, $P_{\phi}$ = 8.47~min, $L_{i} \geq 1.81L_{s}$\\
The average rotational velocities were $\overline{\dot\phi}$ $\sim$ 42.5~$deg$$\cdot$$min^{-1}$ and $\overline{\dot\psi}$ $\sim$ 29.4~$deg{\cdot}min^{-1}$. Moreover, we applied $L_{i}$ = 8~m as the effective diameter of 2012 $\mathrm{TC_4}$. In addition, the moments of inertia of Eq. (A12) were given as follows:
\begin{equation}
\frac{M^2}{2E} = \frac{(nI_{s} + I_{i})}{n+1} \quad or \quad \mbox{ }\frac{(I_{s} + mI_{i})}{m+1},
\end{equation}
where $n$ is an integer from 1 to 9 and $m$ is an integer from 2 to 9. The tumbling status were roughly described by changing the integer $n$ and $m$. The combination of the $n$, $m$, and axial lengths that satisfies the observed velocities $\overline{\dot\phi}$ and $\overline{\dot\psi}$ was also sought. The axial lengths were scanned in steps of 0.1 m for each case. The result was the following combination:\\
$\bullet$Case 6': ($L_{s}$, $L_{i}$, $L_{l}$) = (4.2 m, 8.0 m, 19.2 m), $M^2/2E$ = $(7I_{l} + I_{i})/8$\\
There was no solution that satisfied the observed velocities $\overline{\dot\phi}$ and $\overline{\dot\psi}$ for ``Case 5". 

Finally, the shape and rotational motion models (models 1, 3, 4, and 6) were made for four cases (Cases 1', 3', 4' and 6') using a commercial 3DCG (3-dimensional computer graphic) software, Shade 3D\footnote{Shade 3D is produced by Shade3D Co., Ltd. https://shade3d.jp/en/}, and selecting the plausible models using the artificial lightcurve produced by the shape and rotational motion models. The rendering models of 2012 $\mathrm{TC_4}$  were drawn with the ray-tracing method in the software. The rotational motion was simplified with the fixed rotational velocities of $\overline{\dot\phi}$ and $\overline{\dot\psi}$. The rendering models of 2012 $\mathrm{TC_4}$ were read out in BMP format in steps of 0.1~mins. The artificial lightcurves were produced by the brightness change in the image of each rendering model. The artificial lightcurves are shown in Figure 5. The artificial lightcurves do not rigorously reflect the direction of rotational angular momentum or the detailed topography of 2012 $\mathrm{TC_4}$. However, the shapes of the lightcurves help to narrow down the plausible models. When the lightcurve in Figure 2 is compared to the artificial lightcurves of Figure 5, it can be seen that the artificial lightcurves of models 1 and 6 do not match the observed lightcurve, with respect to the unchanged lightcurve amplitude. On the other hand, it can be seen that the artificial lightcuves of models 3 and 4 created changes in the amplitude, like the observed lightcurve. Therefore, we concluded models 3 and 4 were the plausible models of 2012 $\mathrm{TC_4}$. For any case of models 3 and 4, the shape of 2012 $\mathrm{TC_4}$ is flat and elongated like a pancake. As an example, the shape of model 4 is shown in Figure 6, and the time variation of rotational motion is shown in Figure 7. The rotational motion in model 3 is omitted, since the rotational motion of model 3 is similar to that of model 4. We summarize the shape and rotational motion models of 2012 $\mathrm{TC_4}$ in Table 4. The average $\theta$ value is around 29.0 $deg$ and oscillates within the range of $\pm$0.4 $deg$ in model 3. The average $\dot\phi$ value is around 29.4 $deg{\cdot}min^{-1}$ and oscillates within the range of $\pm$1.5 $deg{\cdot}min^{-1}$ in model 3. The average $\theta$ value is around 48.5 $deg$ and oscillates within the range of $\pm$1.5 $deg$ in model 4. The average $\dot\phi$ value is around 29.4 $deg{\cdot}min^{-1}$ and oscillates within the range of $\pm$2.4 $deg{\cdot}min^{-1}$ in model 4. The $\dot\psi$ is almost constant at around 42.5 $deg{\cdot}min^{-1}$ for both models 3 and 4. The change of $\dot\psi$ is not obvious at the second row of Figure 7. 

\section{Discussion}
We have found that 2012 $\mathrm{TC_4}$ is a tumbling, fast-rotating, and monolithic asteroid. Discoveries of fast-rotating asteroid are increasing with asteroid surveys and follow-up observations (\citealp{Pravec00}; \citealp{Her11}). The number of fast-rotating asteroids is 84 in the Lightcurve Database (LCDB) with``Quality = 3 (denotes a secure result with no ambiguity and full lightcurve coverage)" \citep{Warner09}. Out of these, tumbling asteroids are 2000 WL$_{107}$ \citep{Pravec05}, 2008 $\mathrm{TC_3}$ (\citealp{Betzler09}; \citealp{Scheirich10}), 2004 FH  (LCDB), 2013 SU$_{24}$ (\citealp{Warner14}; \citealp{Benishek14}), 2014 SC$_{324}$ \citep{Warner15}, 2015 VY$_{105}$ \citep{Carbognani16}, 2016 QS$_{11}$ (LCDB), 2018 AJ \citep{Warner18b}, and 2012 $\mathrm{TC_4}$. Moreover, only 2008 $\mathrm{TC_3}$ had been revealed the three axis ratios. This study of 2012 $\mathrm{TC_4}$ is a second sample to the elucidation of three axis ratios. We discuss the formation process of a tumbling and \replaced{monolithic}{fast-rotating} asteroid. The causes of a precession motion were proposed to be impact events with another object, planet encounters, and the decrease of rotational velocity due to the YORP effect \citep{Pravec05}. The precession motion by the planet encounters works  effectively for slow-rotating asteroids \citep{Scheeres04}. Furthermore, fast-rotating asteroids do not start the precession motion due to the decrease of rotational velocity by the YORP effect. Therefore, the cause of precession motion for a tumbling and \replaced{monolithic}{fast-rotating} asteroid is an impact event with another object. Assuming an albedo of 0.2, the diameters of nine tumbling and \replaced{monolithic}{fast-rotating} asteroids, including that of 2012 $\mathrm{TC_4}$, are smaller than 41~m in diameter. The small diameter indicates that the nine tumbling and fast-rotating asteroids would be ejected objects by the impact event rather than the impacted parent objects. \cite{Michi10} pointed out that the axial ratio of the intermediate axis to the long axis of fast-rotating asteroids (diameter $<$ 200~m and rotational period $<$ 1~h) is similar to that of ejecta in laboratory impact experiments and that of boulders on Itokawa and Eros. For example, $L_{i}/L_{l}$, the mean value of axial ratios of boulders larger than 5~m on Itokawa is 0.61 $\pm$ 0.19. Since the lightcurve amplitudes of nine tumbling asteroids are larger than 1.0 mag, the shape of nine tumbling asteroids presumably indicates elongated boulder-like shapes. In particular, the axial ratio $L_{i}/L_{l}$ of 2012 $\mathrm{TC_4}$ is 0.54 in model 3 and 0.56 in model 4, and the axial ratio $L_{i}/L_{l}$ of 2008  $\mathrm{TC_3}$ is 0.54. The NEOs, 2012 $\mathrm{TC_4}$ and 2008  $\mathrm{TC_3}$, will be objects similar to the boulders on Itokawa. Furthermore, we discuss how the impact event happened to fast-rotating asteroids using the axis ratio, $L_{s}/L_{l}$. The collisional destruction process is divided into impact cratering (low impact energy) and catastrophic disruption (high impact energy). Laboratory impact experiments demonstrated that $L_{s}/L_{l}$ of impact cratering fragments is $\sim$0.2, $L_{s}/L_{l}$ of catastrophic disruption fragments is $\sim$0.5, and $L_{s}/L_{l}$ decreases with decreasing impact energy \citep{Michi16}. Numerous impact fragments were generated by the laboratory impact experiments. Despite of the catastrophic disruption, a part of the impact fragments will indicate low $L_{s}/L_{l}$. Thus, the collisional destruction process cannot be immediately concluded from the $L_{s}/L_{l}$ of asteroids. Nonetheless, the axial ratio $L_{s}/L_{l}$ of 2012 $\mathrm{TC_4}$ is 0.42 in model 3 and 0.23 in model 4, and the axial ratio $L_{s}/L_{l}$ of 2008 $\mathrm{TC_3}$ is 0.36.  The NEO 2012 $\mathrm{TC_4}$ could be generated by catastrophic disruption in model 3, and by impact cratering in model 4. The NEO 2008 $\mathrm{TC_3}$ could have experienced the impact energy between models 3 and 4.  

As we discussed above, 2012 $\mathrm{TC_4}$ had possibly experienced an impact event. Here, we estimate the excitation and damping timescales of 2012 $\mathrm{TC_4}$. The excitation timescale, especially, helps to deduce the time of the impact event of 2012 $\mathrm{TC_4}$. An nutation angle ($\theta$ in the Appendix) of asteroids with the LAM increases with dissipating the internal energy. Then, the motion of the asteroid transitions to the SAM via an unstable and temporary rotation mode around the intermediate axis. After the transition to the SAM, the nutation angle decreases with the time, and the SAM transitions to the pure rotation around the short axis, which is in alignment with the principal axis of moment of inertia. We call the transition time from the LAM to the SAM ``excitation timescale", and the transition time from the SAM to the pure rotation ``damping timescale". The excitation and damping timescale (\citealp{Sharma05}, \citealp{Breiter12}) are expressed as
\begin{equation}
T_{s}=D_{s}(h_{1},h_{2})\frac{{\mu}Q}{a^{2}{\rho}{\tilde{\omega}^{3}_{s}}},
\end{equation}
where $D_{s}(h_{1},h_{2})$ is a shape parameter; $\mu$ is the elastic modulus; $Q$ is the quality factor; $\rho$ is the density; $a$ is the half of the long axis length; and $\tilde{\omega}_{s}$ is a representative angular velocity around the focusing principal axis. The quantities for the LAM have the subscript $s = 1$, and those for the SAM have the subscript $s = 3$. The shape parameter $D_{s}(h_{1},h_{2})$ for the LAM and SAM are defined as
\begin{equation}
D_{1}(h_{1},h_{2}) = \left[\frac{h^{2}_{1}(1-h^{2}_{1})(1+h^{2}_{2})}{5(1+h^{2}_{1}h^{2}_{2})}\right]\int_{{\theta}^{0}_{1}}^{{\theta}^{'}_{1}}\frac{sin\theta_{1}cos\theta_{1}}{\Psi_{1}}d\theta_{1},
\end{equation}
and
\begin{equation}
D_{3}(h_{1},h_{2})= -\left[\frac{h^{2}_{1}(1+h^{2}_{1})(1-h^{2}_{2})}{5(1+h^{2}_{1}h^{2}_{2})}\right]\int_{{\theta}^{0}_{3}}^{{\theta}^{'}_{3}}\frac{sin\theta_{3}cos\theta_{3}}{\Psi_{3}}d\theta_{3},
\end{equation}
where $\theta^{0}_{s}$ and $\theta^{'}_{s}$ are the initial and the final maximum wobbling angle, respectively, $h_{1}$ $\equiv$  $L_{i}/L_{l}$, $h_{2}$ $\equiv$ $L_{s}/L_{i}$, $\Psi_{1}$ and $\Psi_{3}$ are dimensionless factor of the energy loss rate \citep{Breiter12}. According to the manner of \citealt{Pravec14}, ${\tilde{\omega}_{1}}$ and  ${\tilde{\omega}_{3}}$ are represented as
\begin{equation}
{\tilde{\omega}_{1}}= \frac{I_{i}}{I_{s}}{\tilde{\omega}_{2}} \equiv \frac{1+h^{2}_{1}h^{2}_{2}}{h^{2}_{1}(1+h^{2}_{2})}{\tilde{\omega}_{2}},
\end{equation}
and
\begin{equation}
{\tilde{\omega}_{3}}= \frac{I_{i}}{I_{l}}{\tilde{\omega}_{2}} \equiv \frac{1+h^{2}_{1}h^{2}_{2}}{1+h^{2}_{1}}{\tilde{\omega}_{2}},
\end{equation}
where $I_{s}$, $I_{i}$, and $I_{l}$ are the moment of inertias defined in the Appendix. When we use ${\tilde{\omega}_{obs}}$ $\equiv$ 2$\pi$/$P_{\phi}$ as a proxy for ${\tilde{\omega}_{2}}$ and $a$ $ \equiv$ $D_{m}/2h_{1}$, where $D_{m}$ is the asteroid mean diameter, the final formulae for the excitation and damping timescales become
\begin{equation}
T_{1} = D_{1}(h_{1},h_{2})\frac{\left(h^{2}_{1}(1+h^{2}_{2})\right)^{3}h^{2}_{1}{\mu}QP_{\phi}^{3}}{(1+h^{2}_{1}h^{2}_{2})^{3}2\pi^{3}{\rho}D^{2}_{m}},
\end{equation}
and
\begin{equation}
T_{3} = D_{3}(h_{1},h_{2})\frac{(1+h^{2}_{1})^{3}h^{2}_{1}{\mu}QP_{\phi}^{3}}{(1+h^{2}_{1}h^{2}_{2})^{3}2\pi^{3}{\rho}D^{2}_{m}}.
\end{equation}
We adopted $D_{m}$ = 8 m, ${\mu}$ = $10^{9}$ Pa, $Q$ = 100, and ${\rho}$ = 3000 kg m$^{-3}$ for 2012 $\mathrm{TC_4}$. The typical timescale from the impact event to the status of model 3 became 3.1 $\times$ 10$^{5}$ yr when the integration interval of $\theta_{1}$ was from 0.1~$deg$ to 29.0~$deg$ in Eq. (17). The status of model 3 transitions to the SAM in the timescale of 2.7 $\times$ 10$^{5}$ yr when the integration interval of $\theta_{1}$ was from 29.0~$deg$ to 89.9~$deg$ in Eq. (17). After the transition to the SAM, the damping timescale, $T_{3}$ became 1.5 $\times$ 10$^{7}$ yr when the integration interval of $\theta_{3}$ was from 89.9~$deg$ to 0.1~$deg$ in Eq. (18). In the same way, the typical timescale from the impact event to the status of model 4 became 3.2 $\times$ 10$^{5}$ yr when the integration interval of $\theta_{1}$ was from 0.1~$deg$ to 48.5~$deg$. The status of model 4 transitions to the SAM in the timescale of 1.8 $\times$ 10$^{5}$ yr when the integration interval of $\theta_{1}$ was from 48.5~$deg$ to 89.9~$deg$. The damping timescale, $T_{3}$ became 3.8 $\times$ 10$^{7}$ yr when the integration interval of $\theta_{3}$ was from 89.9~$deg$ to 0.1~$deg$. On the basis of the excitation and damping timescales, we can make the following scenario of 2012 $\mathrm{TC_4}$. 
\cite{Zap02}, \cite{Morbi03}, and \cite{Granvik17} described that impact events and dynamical mechanisms like the Yarkovsky effect continuously supply asteroids to the transportation resonances in the asteroid Main Belt. If asteroids once move into the transportation resonances, the orbit dynamically evolves to the NEO region in less than a million years \citep{Morbi02}. After the migration to the NEO region, the dynamical lifetime of a 10~m-sized NEO is typically a few million years. 
In the case of 2012 $\mathrm{TC_4}$, its parent object had experienced an impact event in the asteroid Main Belt within $\sim$3 $\times$ 10$^{5}$ yr and the ejected 2012 $\mathrm{TC_4}$ dynamically evolved to the NEO region via the transportation resonances. Even if the derived $\theta$ values was underestimated, the ongoing LAM of 2012 $\mathrm{TC_4}$ is an evidence that the impact even should have happened less than $\sim$6 $\times$ 10$^{5}$ yr ago. The result suggests that  2012 $\mathrm{TC_4}$ should have a fresh surface, since 2012 $\mathrm{TC_4}$ is not exposed to space weathering for more than $\sim$6 $\times$ 10$^{5}$ yr. The motion of 2012 $\mathrm{TC_4}$ will transition to the SAM in $\sim$3 $\times$ 10$^{5}$ yr and then will reach the dynamical lifetime of the 10-m sized NEOs before the damping timescale of tens of million year elapses.

\section{Summary}
We investigated the physical properties of 2012 $\mathrm{TC_4}$ by visible and near-infrared photometry. We succeeded in obtaining unprecedented high time resolution lightcurve with the Tomo-e Gozen camera. The two fast-rotating periods showed that 2012 $\mathrm{TC_4}$ is a tumbling and monolithic asteroid. The observations demonstrated the Tomo-e Gozen camera to be an extremely suitable instrument to observe fast-rotating and fast-moving asteroids. The multiband photometry indicated the taxonomic class of 2012 $\mathrm{TC_4}$ to be an X-type. Assuming the typical albedo of the X-type asteroids, the diameter of 8~m and range of 6-20~m were deduced. Moreover, the shape and rotational motion models of 2012 $\mathrm{TC_4}$ were estimated. The plausible models indicated that 2012 $\mathrm{TC_4}$ has the rotational period of 8.47~min and precession period of 12.25~min with the LAM mode. The three axial lengths were 6.2 $\times$ 8.0 $\times$ 14.9~m or 3.3 $\times$ 8.0 $\times$ 14.3~m. In any model, the shape of 2012 $\mathrm{TC_4}$ is flattened and elongated like a pancake, which suggests that 2012 $\mathrm{TC_4}$ was produced by a past impact event. We also estimated the excitation and damping timescales. The excitation timescale implies that the impact event happened within $\sim$3 $\times$ 10$^{5}$ yr and 2012 $\mathrm{TC_4}$ has a fresh surface that has not been strongly influenced by the space weathering.

This study is a detailed observation of 10~m-sized small NEOs, following the study of 2008 $\mathrm{TC_3}$. Although the impact of a 10~m-sized NEO dose not cause a catastrophic disaster, the impact happens with a high probability from once a century to once in several decades (\citealp{Toricarico17}; \citealp{Trilling17}). It will become a crisis close to the Chelyabinsk meteor event \citep{Popova13}. Furthermore, future space explorations plan to use 10~m-sized NEOs as resources. Thus, clarifying the physical properties of 10~m-sized NEOs is important for both planetary defense and future space exploration.

%\clearpage
\newpage
\appendix
\section{Motion of Force-Free Rigid Body}
The shape of an asteroid is approximated by a triaxial ellipsoid with the axial lengths $L_{s}$, $L_{i}$, and $L_{l}$. The tumbling motion are divided into two classes: the long axis mode (LAM) and short axis mode (SAM). Here, the moment of inertia per unit mass for each axis can be described as 

\begin{equation}
I_{l} = \frac{1}{20}(L_{i}^2 + L_{s}^2),
\end{equation}
\begin{equation}
I_{i} = \frac{1}{20}(L_{l}^2 + L_{s}^2),
\end{equation}
and
\begin{equation}
I_{s} = \frac{1}{20}(L_{l}^2 + L_{i}^2).
\end{equation}
The motion for LAM  can be expressed in terms of the total rotational angular momentum $M$ and total rotational energy $E$ as  

\begin{equation}
I_{l} \leq  \frac{M^2}{2E} < I_{i}. 
\end{equation}
The body approaches pure rotation about the long axis as ${M^2}/{2E}$ approaches $I_{l}$. A new independent variable of time $\tau$ and a constant of the motion $k^2 (\leq 1)$ are defined by 

\begin{equation}
\tau =  t\sqrt{\frac{2E(I_{i} - I_{l})\left( I_{s} - \cfrac{M^2}{2E}\right)}{I_{l}I_{i}I_{s}}},
\end{equation}
and
\begin{equation}
k^2 =  \frac{(I_{s} - I_{i})\left(\cfrac{M^2}{2E} - I_{l}\right)}{(I_{i} - I_{l})\left(I_{s} - \cfrac{M^2}{2E}\right)}.
\end{equation}
The motion of a triaxial ellipsoid can be described as the time-series change of Euler angles $\psi$, $\phi$, and $\theta$. In the case of LAM, $\psi$ is the rotation about the long axis; $\phi$ is the precession about the total rotational angular momentum vector; $\theta$ is the angle between the long axis and total rotational angular momentum vector $M$.  $\psi$, $\theta$ and $\dot\phi$ are
 described as 
 
\begin{equation}
\psi =  atan2\left(\sqrt{\frac{I_{i}}{I_{i} - I_{l}}} {\rm sn}\tau, \sqrt{\frac{I_{s}}{I_{s} - I_{l}}} {\rm cn}\tau\right), 
\end{equation}

\begin{equation}
\theta =  cos^{-1}\left( {\rm dn}\tau\sqrt{\frac{I_{l}\left(I_{s} - \cfrac{M^2}{2E}  \right)}{\cfrac{M^2}{2E} (I_{s} - I_{l})}} \right),
\end{equation}
and
\begin{equation}
\dot\phi =  M\left[\frac{\left( I_{i} - I_{l}\right) + \left(I_{s} -  I_{i} \right){\rm sn}^{2}\tau}  {I_{s}\left(I_{i} - I_{l}\right) + I_{l}\left(I_{s} -   I_{i} \right){\rm sn}^{2}\tau}\right].
\end{equation}
Here, ${\rm sn}\tau$,  ${\rm cn}\tau$ and ${\rm dn}\tau$ are Jacobian elliptic functions. In addition, the following relational expressions are established:
\begin{equation}
P_{\psi} =  4\sqrt{\frac{I_{l}I_{i}I_{s}}{2E(I_{i} - I_{l})\left(I_{s}-\cfrac{M^2}{2E} \right)}}\int_0^\frac{\pi}{2}\frac{du}{\sqrt{1 - k^2sin^2u}},
\end{equation}
and 
\begin{equation}
\frac{P_{\psi}}{P_{\phi}} \geq \sqrt{\frac{(L_{l}^2+L_{i}^2)(L_{l}^2+L_{s}^2)}{(L_{l}^2-L_{i}^2)(L_{l}^2-L_{s}^2)}}-1. 
\end{equation}
The integral part of (A10) is the complete elliptic integral of the first kind.

On the other hand, the motion for SAM  can be expressed as

\begin{equation}
I_{i} <  \frac{M^2}{2E} \leq I_{s}. 
\end{equation}
The body approaches pure rotation about the short axis as ${M^2}/{2E}$ approaches $I_{s}$. A new independent variable of time $\tau$ and a constant of the motion $k^2 (\leq 1)$ are defined by

\begin{equation}
\tau =  t\sqrt{\frac{2E(I_{s} - I_{i})\left( \cfrac{M^2}{2E} - I_{l}\right)}{I_{l}I_{i}I_{s}}},
\end{equation}
and
\begin{equation}
k^2 =  \frac{(I_{i} - I_{l})\left(I_{s} - \cfrac{M^2}{2E} \right)}{(I_{s} - I_{i})\left( \cfrac{M^2}{2E} - I_{l}\right)}.
\end{equation}
In the case of SAM, $\phi$ is the rotation about the short axis; $\psi$ is the oscillation about the long axis; $\theta$ is the angle between the long axis and total rotational angular momentum vector $M$. $\psi$, $\theta$, and $\dot\phi$ are
 described as

\begin{equation}
\psi =  atan2\left(\sqrt{\frac{I_{i}\left(I_{s} - \cfrac{M^2}{2E}  \right)}{I_{s} - I_{i}}} {\rm sn}\tau,  \sqrt{\frac{I_{s}\left(\cfrac{M^2}{2E}  - I_{l},\right)}{I_{s} - I_{l}}}{\rm dn}\tau\right),
\end{equation}

\begin{equation}
\theta =  cos^{-1}\left( {\rm cn}\tau\sqrt{\frac{I_{l}\left(I_{s} - \cfrac{M^2}{2E}  \right)}{\cfrac{M^2}{2E} (I_{s} - I_{l})}} \right),
\end{equation}
and 
\begin{equation}
\dot\phi =  M\left(\frac{\left( \cfrac{M^2}{2E} - I_{l}\right) + \left(I_{s} -   \cfrac{M^2}{2E} \right){\rm sn}^{2}\tau}  {I_{s}\left(\cfrac{M^2}{2E} - I_{l}\right) + I_{l}\left(I_{s} -   \cfrac{M^2}{2E} \right){\rm sn}^{2}\tau}\right).
\end{equation}
In addition, the following relational expressions are established:
\begin{equation}
P_{\psi} =  4\sqrt{\frac{I_{l}I_{i}I_{s}}{2E(I_{s} - I_{i})\left(\cfrac{M^2}{2E} - I_{l}\right)}}\int_0^\frac{\pi}{2}\frac{du}{\sqrt{1 - k^2sin^2u}}.
\end{equation}
The integral part of (A18) is the complete elliptic integral of the first kind. Moreover, the rotational period $P_{\phi}$ has the following relationships with the oscillation period $P_{\psi}$: 

\begin{equation}
\frac{P_{\psi}}{P_{\phi}} \geq \sqrt{\frac{(L_{l}^2+L_{s}^2)(L_{i}^2+L_{s}^2)}{(L_{l}^2-L_{s}^2)(L_{i}^2-L_{s}^2)}}, 
\end{equation}
and 
\begin{equation}
\frac{P_{\psi}}{P_{\phi}} > 1. 
\end{equation}

%\subsection{Figures\label{subsec:figures}}
%If eps file shifts, input following command
%ps2pdf -dEPSCrop -dPDFSETTINGS=/prepress $IN.eps $IN.pdf
%pdf2ps $IN.pdf $IN.eps

%% The "ht!" tells LaTeX to put the figure "here" first, at the "top" next
%% and to override the normal way of calculating a float position

%\newpage

\begin{deluxetable*}{cCCcc}[b!]
\tablecaption{Observation states \label{tab:mathmode}}
\tablecolumns{5}
\tablenum{1}
\tablewidth{0pt}
\tablehead{
\colhead{Observation start  and end time\tablenotemark{a}} &
\colhead{Exp.time} & \colhead{Filter} & \colhead{Observatory} & Average SNR\tablenotemark{b}\\
\colhead{(JD-2458000)} & 
\colhead{(s)} & \colhead{} & \colhead{}
}
\startdata
35.9578326 -- 35.9982278 & 10 &  -- & Kiso (1.05~m) & $\sim$12\\
36.9028528 -- 36.9888035 & 30 & $V$, $R$ & Nayoro (0.4~m) & $V$ and $R$ ($\sim$5)\\
36.90450     -- 36.91530     & 120 & $J$, $H$, $K_{s}$ &Nishi-Harima (2.0~m) & $J$($\sim$120) $H$($\sim$240) $K_{s}$($\sim$160) \\
36.9707319 -- 37.1452383 & 60 & $g'$, $r'$, $i'$, $z'$ & BSGC (1.0~m) & $g'$($\sim$19) $r'$($\sim$30) $i'$($\sim$34) $z'$($\sim$20)\\
37.0073257 -- 37.0808347 & 10|5 & -- & Kiso (1.05~m) & $\sim$26 \\
37.9282507 -- 37.9375688 &  5 & grism & Kiso (1.05~m) & $\sim$8\\
37.9390568 -- 38.0126750 &  2 & -- &  Kiso (1.05~m) & $\sim$66\\
38.0826590 -- 38.1130069 &  6 & -- & Anan (1.13~m) & $\sim$7\\
\enddata
\tablenotetext{a}{Center of exposure time. The time is calibrated light-travel time, with the exception of Nishi-Harima.}
\tablenotetext{b}{SNR of Nishi-Harima is estimated by a image of seven stacked frame.}
\end{deluxetable*}

\begin{deluxetable*}{cccc}[b!]
\tablecaption{Status of 2012 $\mathrm{TC_4}$ each day \label{tab:mathmode}}
\tablecolumns{4}
\tablenum{2}
\tablewidth{0pt}
\tablehead{
\colhead{Year/mon/day} &
\colhead{$\Delta$\tablenotemark{a}} &
\colhead{$\alpha$\tablenotemark{b}} & \colhead{Sky motion} \\
\colhead{(UT)} & \colhead{(AU)} &
\colhead{($^\circ$)} & \colhead{$\arcsec$/min} 
}
\startdata
2017/10/9.4578 -- 9.4998 & 0.011 -- 0.010 & 31.4 -- 31.5 &  4.16 -- 4.54\\
2017/10/10.4029 --  10.5808 & 0.007 -- 0.0064 & 33.3 -- 34.1  & 6.77 -- 9.36  \\
2017/10/11.4283 -- 11.6130     &  0.0032 -- 0.0025      & 38.0 -- 40.7 & 28.17 --  43.24 \\
\enddata
\tablenotetext{a}{2012 $\mathrm{TC_4}$ to observer distance.}
\tablenotemark{b}{Phase angle (Sun-2012 $\mathrm{TC_4}$-observer).}
\end{deluxetable*}

\begin{deluxetable*}{cc}[b!]
\tablecaption{Color indexes of 2012 $\mathrm{TC_4}$. \label{tab:mathmode}}
\tablecolumns{2}
\tablenum{3}
\tablewidth{0pt}
\tablehead{
\colhead{Column} &
\colhead{Values} \
}
\startdata
$g - r'$& 0.479 ${\pm}$ 0.031 \\
$r' - i'$& 0.187 ${\pm}$ 0.023 \\
$i' - z'$& 0.035 ${\pm}$ 0.036 \\
$J - H$& 0.226 ${\pm}$ 0.041 \\
$H - K_{s}$& 0.034 ${\pm}$ 0.045 \\
\enddata
\end{deluxetable*}

\begin{deluxetable*}{cc}[b!]
\tablecaption{Shape and rotational motion models of 2012 $\mathrm{TC_4}$. \label{tab:mathmode}}
\tablecolumns{2}
\tablenum{4}
\tablewidth{0pt}
\tablehead{
\colhead{Column} &
\colhead{Values} \
}
\startdata
$L_{s}$& 6.2~$m$ (model 3), 3.3~$m$ (model 4) \\
$L_{i}$& 8.0~$m$ (model 3 and model 4) \\
$L_{l}$& 14.9~$m$ (model 3), 14.3~$m$ (model 4) \\
$P_{\psi}$ & $8.47$ ${\pm}$ 0.01 $min$ \\ 
$P_{\phi}$ & $12.25$ ${\pm}$ 0.01 $min$\\
$\overline{\theta}$ & 29.0~$deg$ (model 3), 48.5~$deg$ (model4)\\
$\overline{\dot\phi}$ & 29.4~$deg{\cdot}min^{-1}$ (model 3 and 4)\\
 $\overline{\dot\psi}$ & 42.5~$deg{\cdot}min^{-1}$ (model 3 and 4)\\
\enddata
\end{deluxetable*}

\clearpage
\begin{figure}[ht!]
\epsscale{0.5}
\plotone{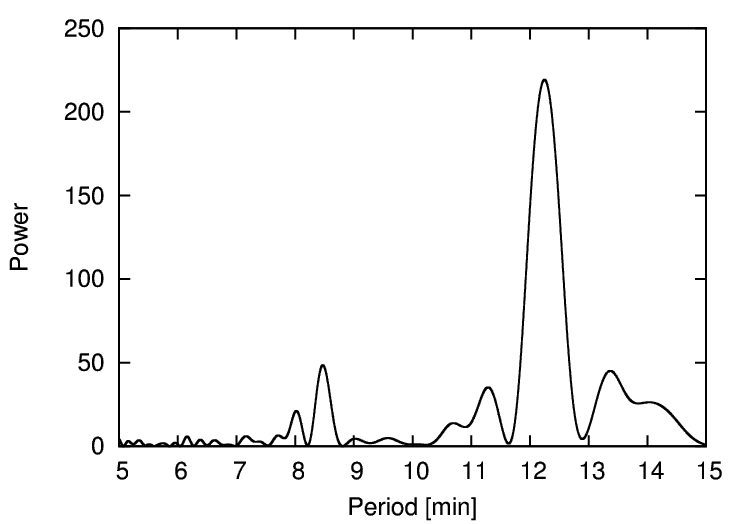}
\caption{Power spectrum for the sidereal rotational period of 2012 $\mathrm{TC_4}$, assuming the double-peak lightcurve. The calculation is carried out by the data obtained on October 10, 2017, at Kiso Observatory. }
%The data comes from Table \ref{tab:table}.\label{fig:general}}
\end{figure}

\begin{figure}[h]
\epsscale{1.2}
\plotone{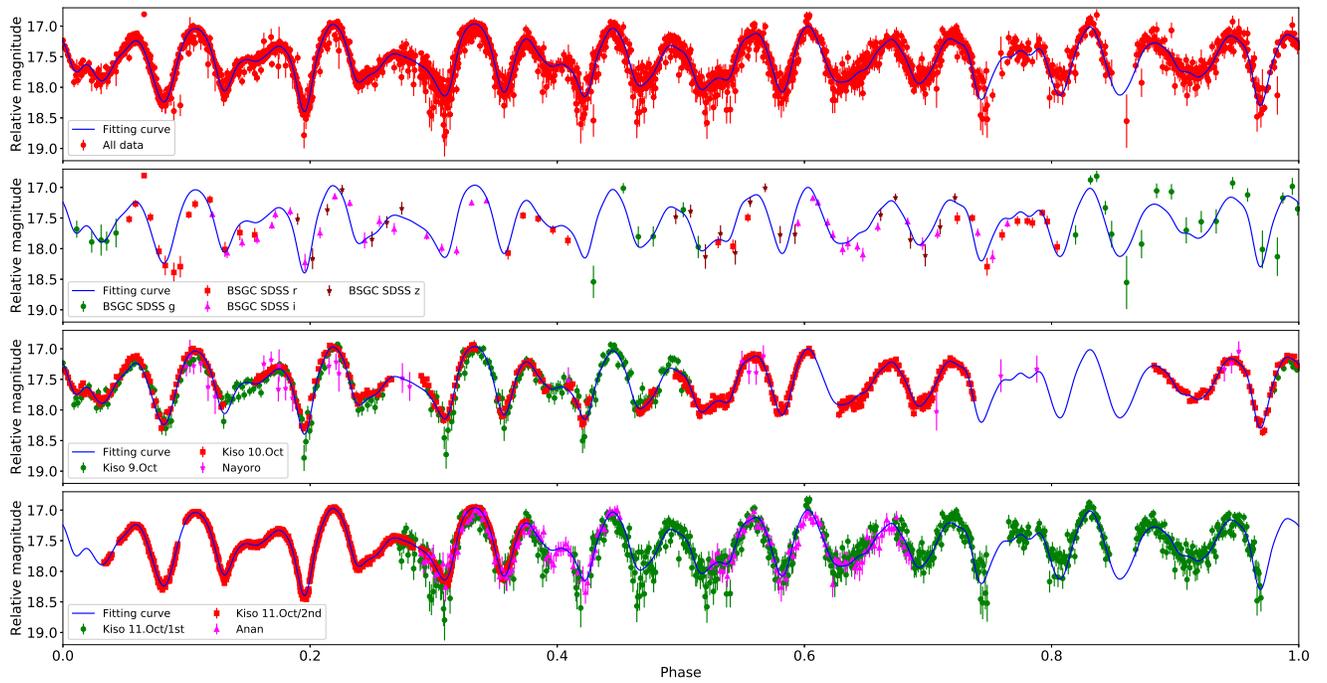}
\caption{Lightcurve of 2012 $\mathrm{TC_4}$. As a result of motion analysis, we deduce the rotational period is 8.47 $\pm$ 0.01 min and the precession period is 12.25 $\pm$ 0.01 min. The phase in the figure indicates the complex period of 110.18~min. The relative magnitude corresponds to the apparent magnitude on October 10, 2017, at Kiso Observatory. Although the apparent magnitude is estimated by comparing the SDSS $g$' magnitude of reference stars, the Tomo-e Gozen is not equipped with the same filter. Therefore, the apparent magnitude is a relative magnitude in a precise sense. (Top) All data and the fitting curve. (2nd row) The timing of the multiband photometry on BSGC. The offset magnitude to the $C_{0}$ for $g'$, $r'$, $i'$, and $z'$ are -0.03 mag, 0.449 mag, 0.636 mag, and 0.674 mag, respectively. }(3rd row) The data on October 9 and 10, 2017,  at Kiso Observatory. (Bottom) The data on October 11, 2017, at Kiso Observatory and Anan Science Center. The graph legend of ``Kiso 11.Oct/1st" is the photometry of the zeroth-order data of grism spectroscopy. The graph legend of ``Kiso 11.Oct/2nd" shows that the Tomo-e Gozen can obtain a precise and high time resolution lightcurve.
%The data comes from Table \ref{tab:table}.\label{fig:general}}
\end{figure}
%\subsection{General figures\label{subsec:general}}

\begin{figure}[ht!]
\epsscale{1}
\plotone{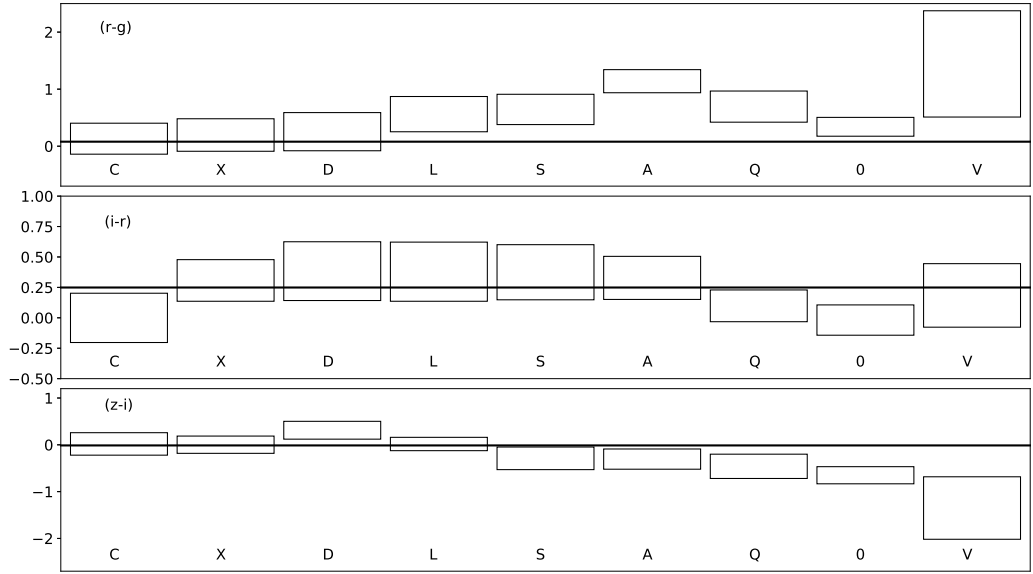}
\caption{Reflectance color gradients of 2012 $\mathrm{TC_4}$ and asteroids of major taxonomic classes. The rectangles indicate the range of reflectance color gradients of C, X, D, L, S, A, Q, O, and V-type asteroids in the SDSS Moving Object Catalog (SDSS-MOC).  The top, middle and bottom figures correspond to the $\gamma_{g}$, $\gamma_{r}$, and $\gamma_{i}$, respectively. The thick horizontal lines are the average reflectance color gradients of 2012 $\mathrm{TC_4}$. The reflectance color gradients of 2012 $\mathrm{TC_4}$ are consistent with the range of X-type asteroids. }
%The data comes from Table \ref{tab:table}.\label{fig:general}}
\end{figure}

\begin{figure}[ht!]
\epsscale{0.5}
\plotone{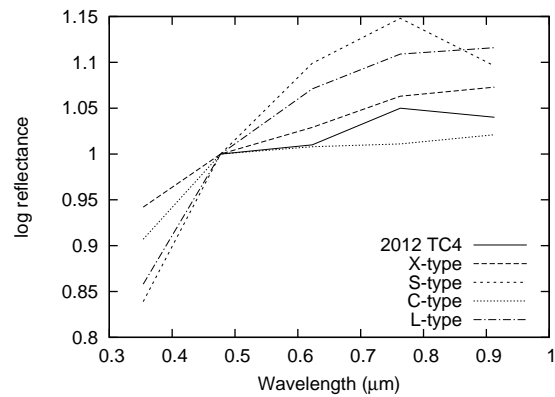}
\caption{ The log reflectance spectra of 2012 $\mathrm{TC_4}$ and the asteroids of the X-, S-, C-, and L-types. The data in $u'$ filter is not obtained for 2012 $\mathrm{TC_4}$.}
%The data comes from Table \ref{tab:table}.\label{fig:general}}
\end{figure}

\begin{figure}[h]
\epsscale{1.2}
\plotone{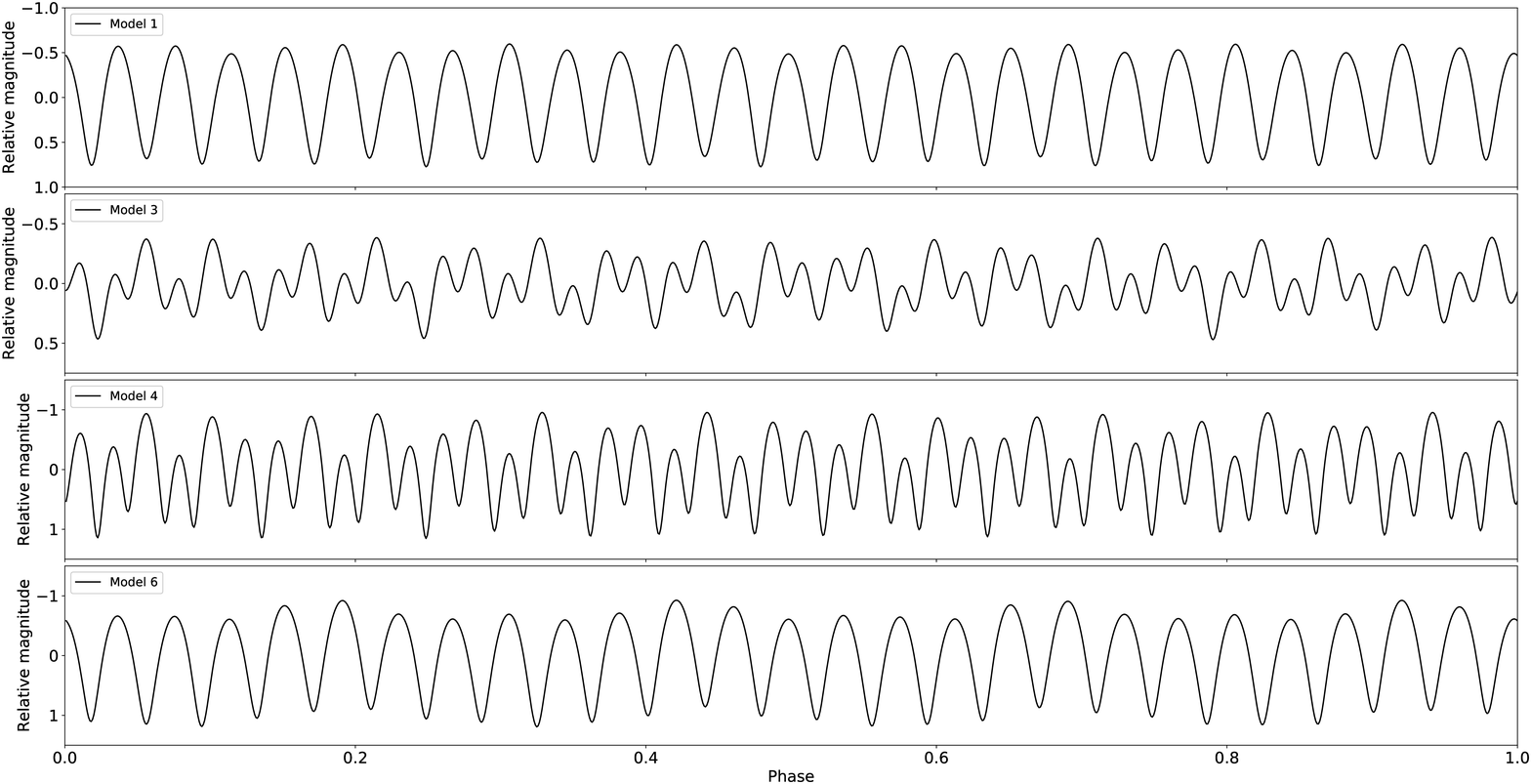}
\caption{Artificial lightcurves of 2012 $\mathrm{TC_4}$ in models 1, 3, 4, and 6.}
%The data comes from Table \ref{tab:table}.\label{fig:general}}
\end{figure}

%\newpage

\begin{figure*}
\gridline{\fig{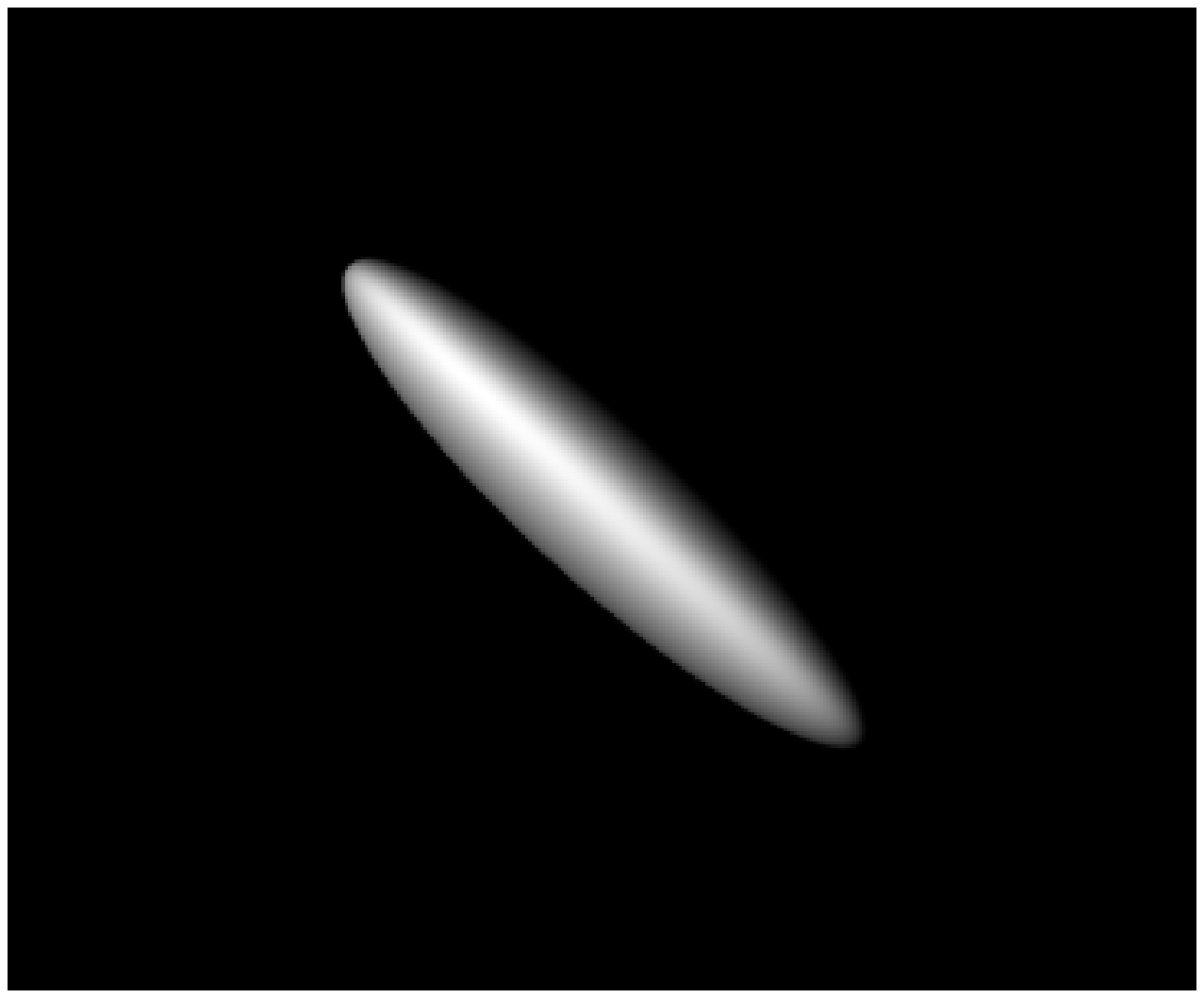}{0.3\textwidth}{(Left)}
          \fig{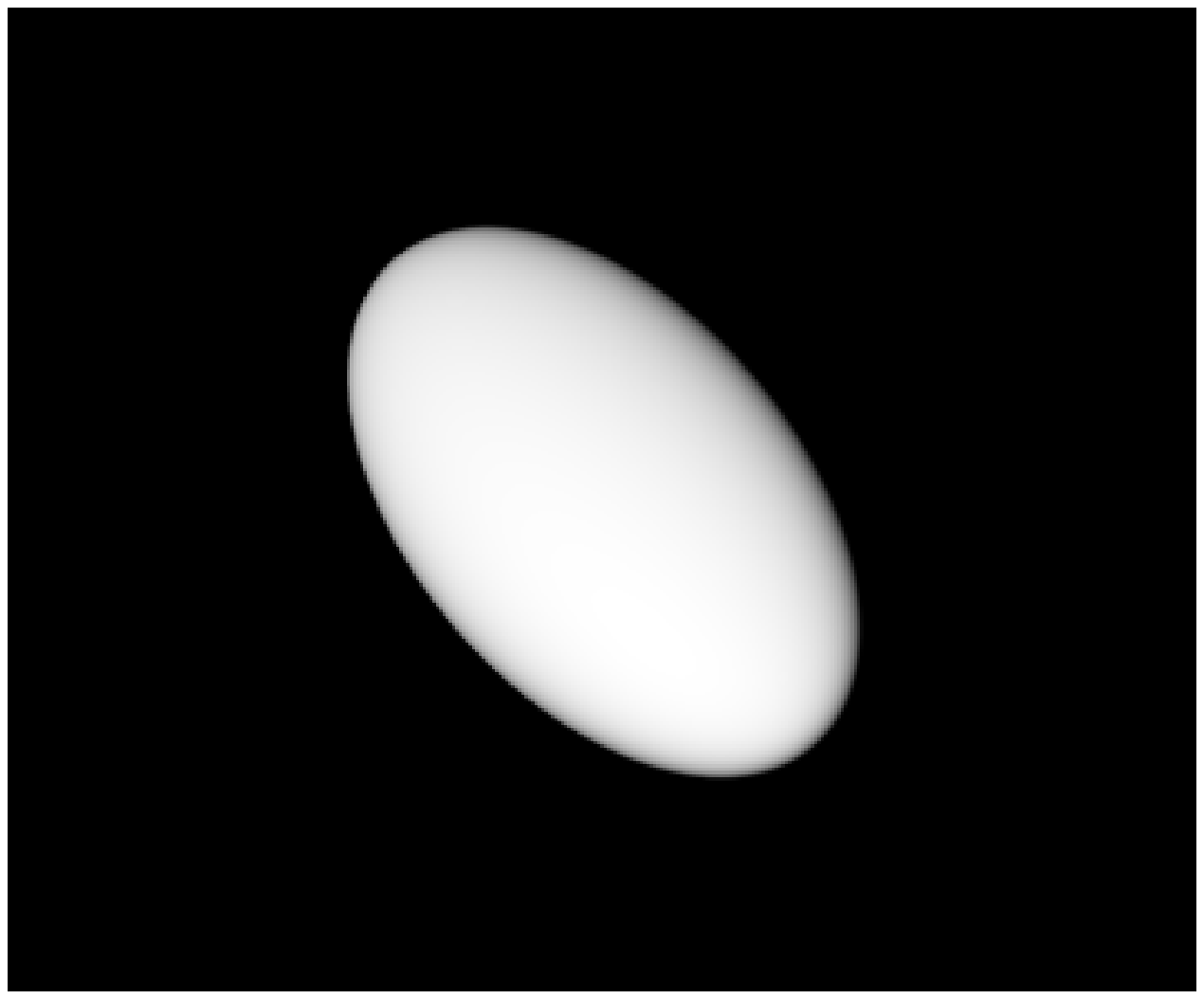}{0.3\textwidth}{(Center)}
          \fig{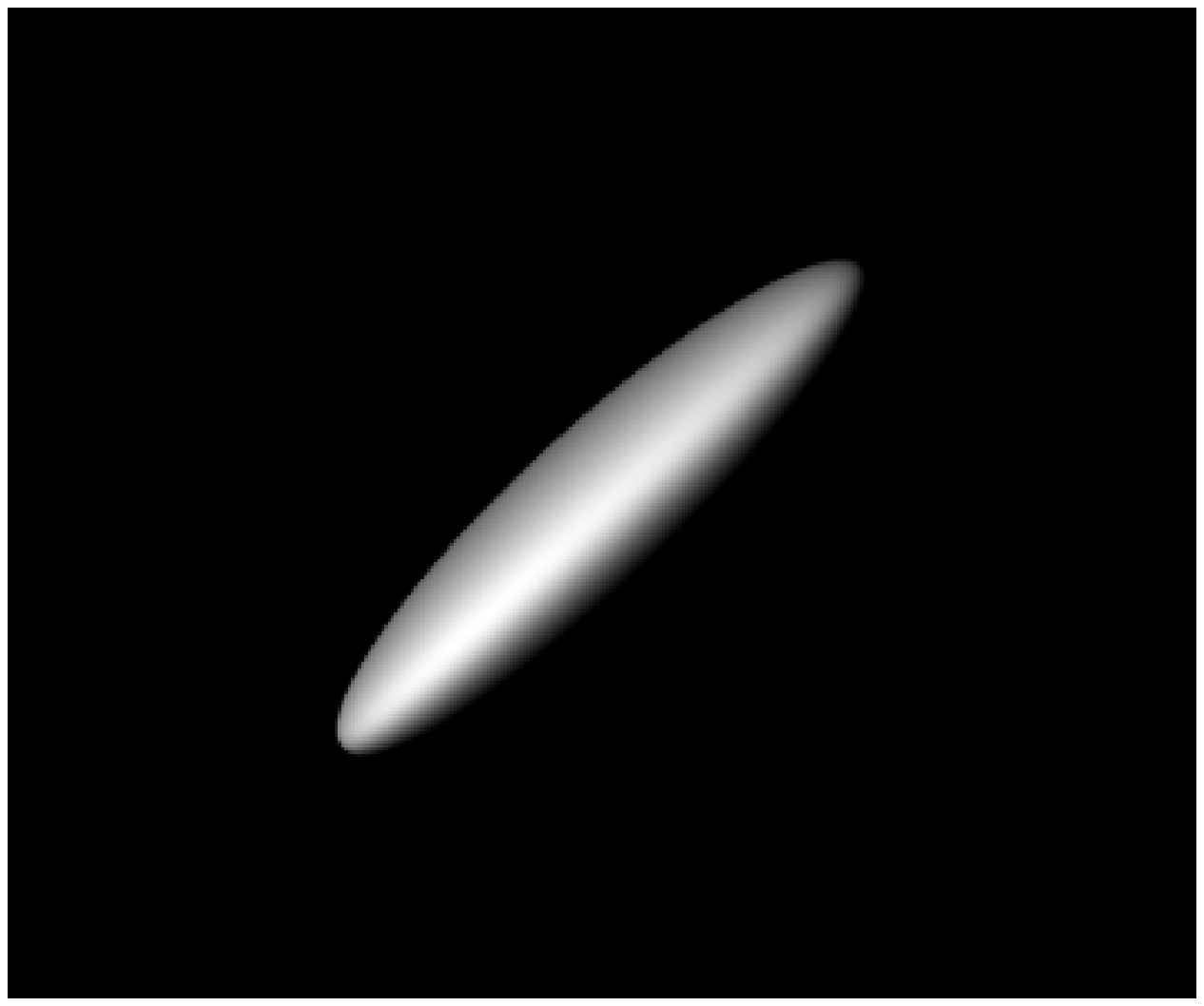}{0.3\textwidth}{(Right)}
          }

\caption{Shape of 2012 $\mathrm{TC_4}$ for a model 4. An observer locates in the direction of the intermediate axis in the left figure. (Left) A view on phase 0 in Figure 5. (Center) A view on phase $\sim$ 0.1 in Figure 5. (Right) A view on phase $\sim$ 0.5 in Figure 5. An animated version of this figure is available in $http://www.spaceguard.or.jp/RSGC/TC4/TC4\_LAM\_4.mp4$\label{fig:pyramid}}
\end{figure*}

\begin{figure}[h]
\epsscale{1}
\plotone{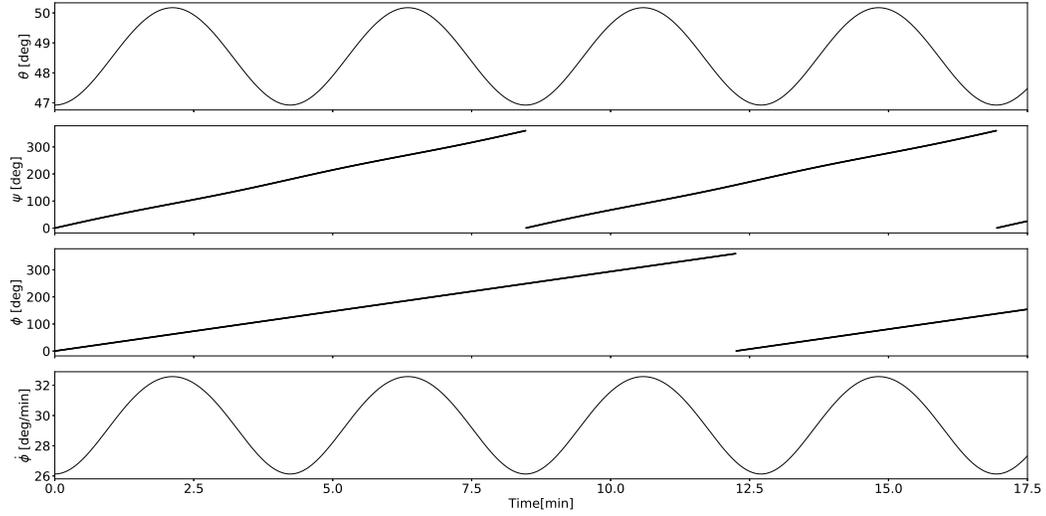}
\caption{The Euler angle of $\theta$, $\phi$, and $\psi$ as a function of time for a model 4. This figure style is developed by \cite{Samarasinha91}. The angle $\theta$ is the angle between the long axis and total rotational angular momentum vector $\bm{M}$. The  angle $\phi$ and $\psi$ measures the amount of precession executed by the long axis around $M$ and amount of rotation around the long axis itself. For model 4, the axial lengths of 3.3 $\times$ 8.0 $\times$ 14.3~m were used with $M^2/2E$ = $(2I_{l}+I_{i})/3$. The nutation period, $P_{\theta}$, is exactly half the rotational period, $P_{\psi}$. The variation in the angular velocity, $\dot\psi$, is undetectable in plots of $\psi$ vs time because the amplitude of variation is negligible. The angle $\phi$ is described based on the constant of $\overline{\dot\phi}$ $\sim$ 29.$^{\circ}$4 minutes$^{-1}$.}
%The data comes from Table \ref{tab:table}.\label{fig:general}}
\end{figure}

\clearpage

%% If you wish to include an acknowledgments section in your paper,
%% separate it off from the body of the text using the \acknowledgments
%% command.
\acknowledgments
This research is supported in part by Japan Society for the Promotion of Science (JSPS) Grants-in-Aid for Scientific Research (KAKENHI) Grant Number 16K05310, JP18H01261, JP26247074, JP16H02158, JP16H06341, JP2905, 18H04575, JP18H01272, JP18K13599, and JSPS Program for Advancing Strategic International Networks to Accelerate the Circulation of Talented Researchers Grant Number JR2603. This research is also supported in part by Japan Science and Technology Agency (JST) Precursory Research for Embryonic Science and Technology (PRESTO), Research Center for the Early Universe (RESCEU), School of Science, the University of Tokyo, and the Optical and Near-infrared Astronomy Inter-University Cooperation Program.

\end{document}